# Observation of acoustically induced transparency for γ-ray photons

Y. V. Radeonychev[1,2], I. R. Khairulin[1], F.G. Vagizov[3,4,2] and Olga Kocharovskaya[4]

[1] *Institute of Applied Physics of the Russian Academy of Sciences,*
*46 Ulyanov Street, Nizhny Novgorod, 603950, Russia,*
[2] *Kazan Physical-Technical Institute, Russian Academy of Sciences,*
*10/7 Sibirsky Trakt, Kazan 420029 Russia*
[3] *Kazan Federal University, 18 Kremlyovskaya Street, Kazan 420008, Russia,*
[4] *Department of Physics and Astronomy and Institute for Quantum Studies and Engineering,*
*Texas A&M University, College Station, TX 77843-4242, USA.*

We report an observation of a 148-fold suppression of resonant absorption of 14.4 keV photons from exp(-5.2) to exp(-0.2) with preservation of their spectral and temporal characteristics in an ensemble of the resonant two-level $^{57}$Fe nuclei at room temperature. The transparency was induced via collective acoustic oscillations of nuclei. The proposed technique allows extending the concept of induced optical transparency to a hard x-ray/γ-ray range and paves the way for acoustically controllable interface between x-ray/γ-ray photons and nuclear ensembles, advancing the field of x-ray/γ-ray quantum optics.

The induced transparency of opaque medium for resonant electromagnetic radiation is a powerful tool for manipulating the field-matter interaction. The study of phenomena associated with transparency induced in natural and artificial quantum and classical systems for resonant electromagnetic radiation in a wide spectral range from microwaves to γ-rays, as well as their applications, is an extremely broad area of research [1-12], extended also to acoustic waves [13,14]. Self-induced transparency [1], transparency via Autler-Townes splitting (ATS) [2,3], and electromagnetically induced transparency (EIT) [4,5] including their analogs in various quantum and classical systems [6-11], are just several examples of numerous techniques for suppression of resonant absorption. Important applications of induced optical transparency such as high refractive index and nonlinearities at the few-photon level, slow and stored light, optical quantum information processing, etc. stimulate a development of similar techniques in the hard x-ray/γ-ray domain. High-energy 10–100-keV photons can be easier detected and tighter focused than optical photons [15], while corresponding resonant recoilless nuclear transitions have orders of magnitude narrower linewidths at room temperature than optical atomic transitions at a comparable solid density [12,16]. These features are promising for realization of very compact and efficient interfaces between single hard x-ray/γ-ray photons and nuclear ensembles. However, the common tools for controlling quantum optical interfaces, such as intense spectrally narrow coherent sources and high-finesse cavities are still unavailable in hard x-ray/γ-ray range, preventing from a direct realization of the basic optical transparency techniques such as EIT and ATS-transparency, for high-energy photons. Several different techniques to control resonant interaction between hard x-ray/γ-ray photons and nuclear ensembles were developed, based on variation of hyperfine or external magnetic field [10,17,18], mechanical displacement (periodic or non-periodic) of an absorber or source with respect to each other including acoustic vibration [12,16,19-27], and placing nuclei into a spatial sandwich-like nano-structure [11]. The 25% reduction in absorption of 14.4-keV photons was observed via anti-crossing of the upper energy sublevels of $^{57}$Fe nuclei in a crystal of FeCO$_3$ taking place at 30 K [10]. The 4.5-fold suppression of 14.4-keV collective coherent emission from two $^{57}$Fe layers imbedded into a specifically designed planar waveguide was reported in [11]. It was observed that acoustic vibration of nuclear absorber leads to appearance of sidebands in the Mossbauer absorption spectra corresponding to a decrease in its resonant opacity [24,25]. In the case of piston-like absorber vibration (synchronous nuclear oscillations), the efficient resonant transformation of the incident quasi-monochromatic radiation into sidebands greatly enhanced its transmittance to 70% demonstrated in [12] and over 80% observed in [20]. It was shown that synchronization of the emerged sidebands can result in a strong amplitude modulation of the transmitted resonant field corresponding to ultra-short near-bandwidth-limited γ-ray pulses [12,21,22,26,27]. Similar effects of the resonant sideband generation and extremely short pulse formation can appear when VUV or XUV quasi-monochromatic radiation propagates through an atomic medium which resonance transition frequency is periodically modulated by a strong infrared or optical field via Stark effect [21,28-30].

However, the possibility of complete transmission of hard x-ray/γ-ray photon through an optically thick resonant nuclear absorber without any alterations of its spectral profile and the temporal waveform has never been demonstrated so far. In this work we show that a proper acoustical vibration of an absorbing medium can provide such full transparency completely eliminating the resonant interaction of γ-ray photons with an optically thick nuclear absorber. We determine the conditions for such acoustically induced transparency (AIT) and demonstrate this technique experimentally for the 14.4-keV recoilless individual photons propagating through the $^{57}$Fe resonant nuclear absorber (Fig.1). We show that the effect of AIT is robust, i.e., sufficiently high degree of transparency may be achieved even in the case of a substantial deviation from the AIT conditions.

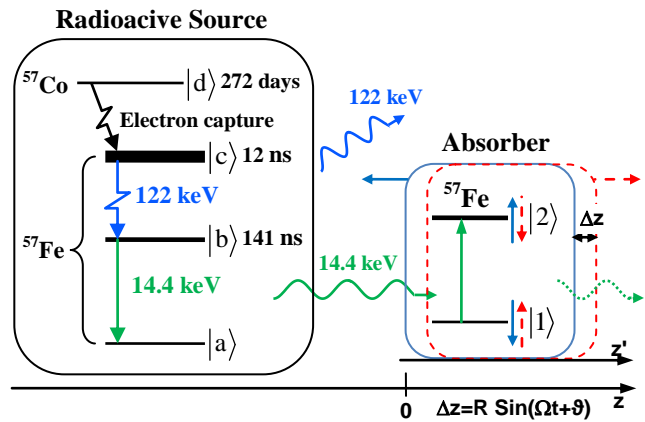

Fig.1. (Color online) **Energy scheme of γ-ray photon source and vibrating absorber.** The $^{57}$Co nuclei (left side) in the state |d⟩ decay to $^{57}$Fe nuclei in the excited state |c⟩ (with half-life of $T_{1/2} \approx 272$ days), followed by cascade decay: |c⟩→|b⟩ (with decay time $T \approx 12$ ns) and |b⟩→|a⟩ (with decay time $T_S \approx 141$ ns) with emission of 122-keV and 14.4-keV photons (shown by blue and green lines), respectively. Depolarized recoilless 14.4-keV photons (λ≈0.86Å) resonantly interact with transition |1⟩→|2⟩ of $^{57}$Fe nuclei when propagating through the single-line $^{57}$Fe absorber (right side). They are resonantly absorbed in motionless absorber. Harmonic vibration of the absorber along the photon propagation direction $z$ (blue and red horizontal arrows) leads to periodic variation in its |1⟩→|2⟩ transition frequency (blue and red vertical arrows) in the laboratory reference frame due to the Doppler effect, which modifies the interaction of photons with absorber.

The physical origin of AIT can be easily understood both in the laboratory reference frame and in the reference frame of the vibrating absorber (Fig.2).

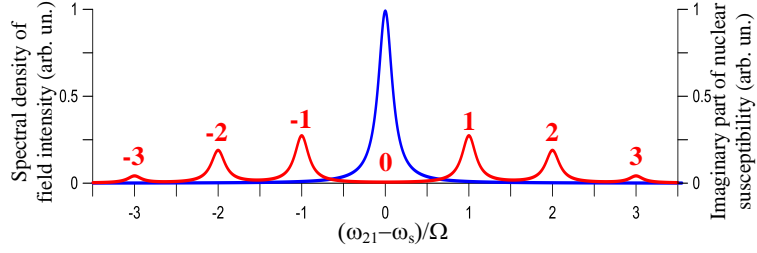

Fig.2. (Color online) **The physical origin of AIT.** In the laboratory reference frame blue curve is the Lorentz spectral profile of the incident field with half-linewidth $\gamma_S$ and red curve is the Lorentz spectral line of the absorber $|1\rangle \rightarrow |2\rangle$ transition (Fig.1) with half-linewidth $\gamma_{21}=\gamma_S$, harmonically vibrating with circular frequency $\Omega/\gamma_{21}=10$ and amplitude $R=R_1$. Equivalently, in the vibrating reference frame blue curve is the spectral line of the same absorber and red curve is the spectral profile of the same field, "seen" by oscillating nuclei [see equations (S7),(S8),(S22),(S27) in Supplemental Material].

In the vibrating reference frame, the spectrum of the incident single-photon wave packet is "seen" by nuclei as a comb of equidistant spectral components separated by the vibration frequency due to the Doppler effect (Fig.2 red line, also see Supplemental Material) [12,22,26,27]. Amplitudes of the spectral components are proportional to Bessel functions of the first kind [12,22,26,27], $J_n(2\pi R/\lambda)$, where $n$ is the number of spectral component, $R$ is the vibration amplitude, and $\lambda$ is the photon wavelength. If the vibration amplitude takes the value

$$R = R_i, \text{ where } R_1 \approx 0.38\lambda, \ R_2 \approx 0.88\lambda, \ R_3 \approx 1.37\lambda, ..., \qquad (1)$$

then $J_0(2\pi R_i/\lambda) \approx 0$, i.e., the resonant (zero) spectral component vanishes and all the photon energy is spread among the sidebands (Fig.2). In the ideal case of monochromatic weak field and infinitely narrow spectral line of the absorber, the spectral sidebands are out of resonance and propagate through the medium without interaction with nuclei. In other words, the absorber is completely transparent for radiation.

Since the spectral lines of both the source and absorber are broadened, the sidebands of the field in the vibrating reference frame overlap with the absorber spectral line (Fig.2). This results in spectrally selective absorption and dispersion of the propagating field, which cannot be neglected for optically deep absorber. Hence, in order to achieve transparency for radiation with half-linewidth $\gamma_S$ in absorber with half-linewidth $\gamma_{21}$ and optical depth $T_M > 1$ (corresponding to $e^{-T_M} < 0.37$), vibrating with circular frequency $\Omega$, two additional conditions should be met,

$$\gamma_S/\Omega \ll 1, \qquad (2a)$$
$$T_M \gamma_{21}/(2\Omega) \ll 1. \qquad (2b)$$

If conditions (1) and (2) are fulfilled, then (i) the resonant spectral component of the field vanishes (Fig.2), and (ii) the sidebands are located so far from the absorber spectral line that both the resonant absorption and dispersion are negligible over the entire propagation length. Hence, neither spectral nor temporal characteristics of the incident field are changed, which is equivalent to complete transparency of the resonant medium. It is worth noting that conditions (1),(2) do not imply the determined initial phase of vibration. This means that AIT can be implemented not only by absorber vibration but also by an acoustic wave or even random oscillations of nuclei with the same amplitude and frequency (also see Supplemental Material).

Equivalently, in the laboratory reference frame, the spectral line of the absorber under AIT-conditions (1),(2) is "seen" by photons as a spectral comb containing only the sidebands but no resonant component (Fig.2), resulting in negligible interaction of photons with nuclei [see equations (S14)-(S15),(S19),(S23)-(S28) in Supplemental Material].

The spectral window of AIT in Fig.2 appears due to a splitting of the absorber spectral line into sidebands having the linewidths of the parent line, similar to the transparency induced by the ATS [2,3] (see also Supplemental Material).

Experimental demonstration of AIT was implemented with the radioactive source $^{57}$Co:Rh and stainless steel foil as the absorber with the optical depth (Mössbauer thickness) $T_M \approx 5.2$, vibrating with frequency $\Omega/(2\pi) \approx 9.87$ MHz (Fig.3-5). The half-linewidths of the source and absorber were $\gamma_S/(2\pi) \approx 0.68$ MHz and $\gamma_{21}/(2\pi) \approx 0.85$ MHz (see Supplemental Material). Hence, conditions (2) were not perfectly satisfied since $\gamma_S/\Omega \approx 0.07$ and $T_M \gamma_{21}/(2\Omega) \approx 0.2$. In order to meet condition (1), the vibration amplitude was adjusted to value $R_1 \approx 0.33$Å and monitored via the Mössbauer transmission spectrum of the vibrating absorber (Fig.3a,b).

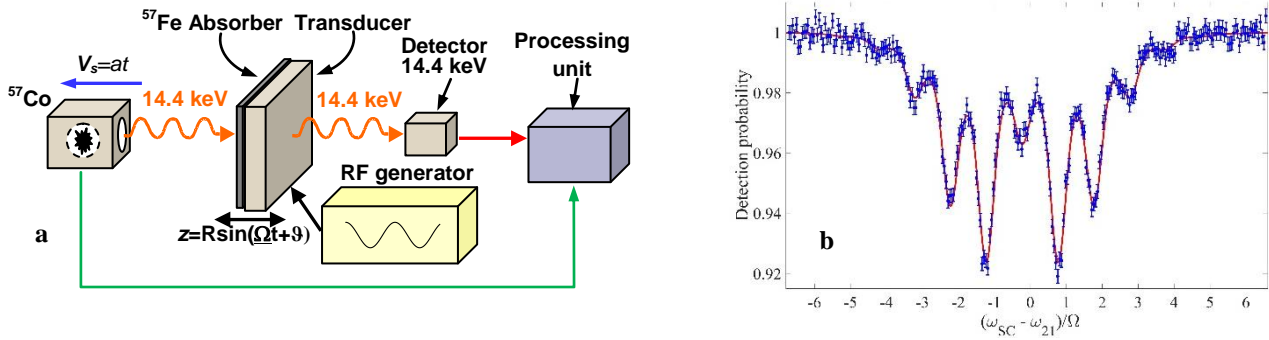

Fig.3. (Color online) **Absorber transmission spectrum.** (a): Generic scheme of the experimental setup. The $^{57}$Co:Rh source is moved with constant acceleration along the photon propagation direction. The absorber is 25µm thickness stainless steel foil of Fe:Cr:Ni; 70:19:11 wt% with natural abundance (~2%) of $^{57}$Fe corresponding to optical depth $T_M \approx 5.2$. The foil is glued on the sinusoidally vibrated polyvinylidene fluoride piezo-electric transducer.
(b): Mössbauer transmission spectrum, i.e., the dependence of probability of detecting the transmitted 14.4-keV photons per unit time on the detuning between the variable central frequency of the source spectral line, $\omega_{SC}$, and the fixed central frequency of the absorber spectral line, $\omega_{21}$, normalized by the transmittance far-from resonance. Blue dots are the measured values with error bars estimated from counting statistics as $\sqrt{N(\omega)}/N_m$, where $N(\omega)$ is the number of counts for frequency ω and $N_m=N(\omega=\pm\infty)$ is the maximum of function $N(\omega)$. Red fitting curve is plotted according to formulas (S35)-(S36) in Supplemental Material.

The transmission spectrum in Fig.3b qualitatively corresponds to the spectral line of the vibrating absorber in Fig.2 (red curve) with two major differences. The first one is the larger broadening of the measured spectral components caused by a number of factors (see Supplemental Material). The second one is the non-vanishing resonant (central) component (unlike red curve in Fig.2). This is due to slightly different vibration amplitudes of different parts of the foil within the beam spot (see Supplemental Material). Hence condition (1) was not perfectly satisfied too. Nevertheless, a substantial suppression of the resonant spectral component compared to the sidebands shows that the most probable vibration amplitude was 0.33Å. Therefore, high transparency of the absorber for 14.4-keV photons could be expected. This was verified via measuring the spectrum and waveform of the transmitted photons.

The measured spectrum of the resonant 14.4-keV single-photon wave packet behind the vibrating absorber in comparison with the cases of in-resonance and far-off-resonance motionless absorber and the corresponding experimental setup are shown in Fig.4.

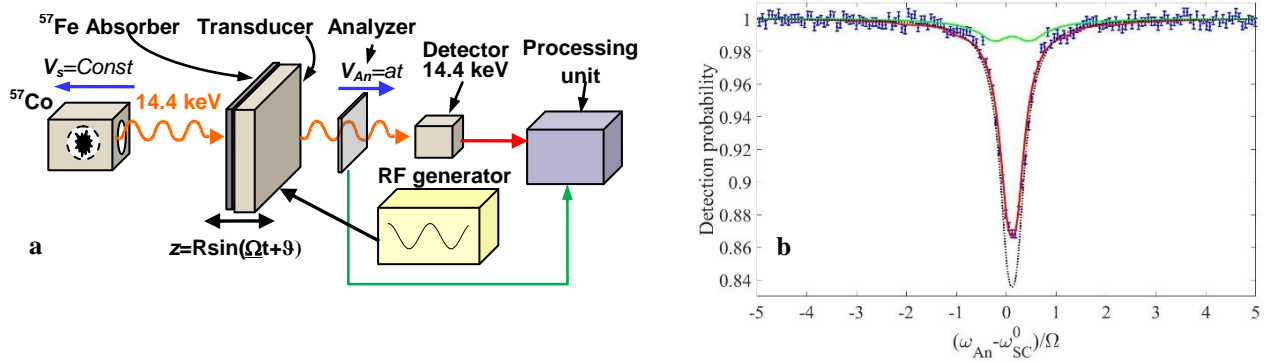

Fig.4. (Color online) **Transmitted photon spectrum.** (a): Generic scheme of the experimental setup. Unlike Fig.3, the source is moved at a constant velocity $v_S$ compensating the isomer shift, so that its resonant frequency $\omega_{SC}^0$ coincides with the absorber resonant frequency. Spectrum of the transmitted radiation is measured by analyzer which is stainless steel foil with the same characteristics as characteristics of the absorber. Analyzer is moved with constant acceleration such that the moment of detection of 14.4-keV photon defines central frequency of analyzer spectral line, $\omega_{An}$ (see Supplemental Material).
(b): Probability of detection of 14.4-keV photon per unit time versus the detuning between variable resonance frequency of analyzer, $\omega_{An}$, and resonance frequency of the source, $\omega_{SC}^0$. Blue dots with error bars are the measured values. Red line is theoretical fitting compared to the case of motionless absorber tuned to resonance (solid green line) or far-from resonance (dotted black line). Theoretical curves are plotted according to formula (S37) in Supplemental Material.

The striking difference between red and green curves in Fig.4b proves that despite the non-perfect implementation of the AIT conditions (1),(2), the resonant absorption is greatly reduced in the vibration absorber as compared to the absorber at rest. Achieved efficiency of AIT evaluated as the ratio between the amplitudes of solid red and dotted black curves, is 82% corresponding to 26 times reduction of the effective optical depth from 5.2 of the motionless absorber to 0.2 in the implemented case of AIT. Similar to the spectral line of the incident photons (black dotted curve in Fig.4b), red line in Fig.4b, has nearly Lorentz profile except of barely noticeable sidebands appeared due to incomplete fulfillment of AIT conditions (1),(2).

The waveform of the transmitted photons (the time-dependence of the photon detection probability proportional to intensity of the single-photon wave packet) was measured via the time-delayed coincidence technique of Mössbauer time-domain spectroscopy (Fig.5a).

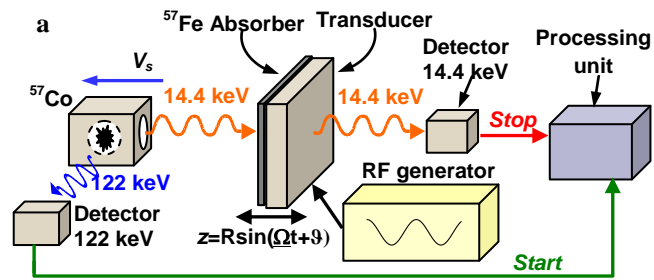

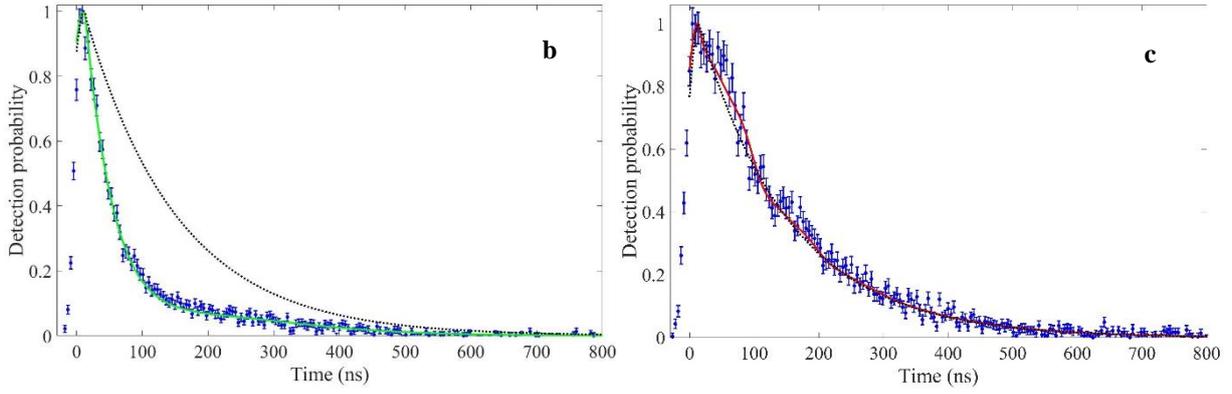

Fig.5. (Color online) **Transmitted photon waveforms.** (a) Generic scheme of the time-delayed coincidence counting sequential 122-keV and 14.4 keV photons. The interval between "Start" and "Stop" signals measured many times gives the coincidence count rate as a function of delay, proportional to the time-dependence of photon detection probability or, equivalently, to the single-photon waveform. The source is moved with constant velocity, $v_S$, to match resonant frequencies of the source and absorber.
(b),(c): The measured waveforms of 14.4-keV photons (blue dots with error bars) transmitted through the motionless resonant absorber (b) and through the vibrating absorber (c) in comparison with the waveform of photon passed through far-off-resonant motionless absorber (black dotted lines). Red, green, and black curves are plotted according to formula (S38) in Supplemental Material. The waveforms in red, green, and black correspond to spectral contours of the same color in Fig.4b.

In the case of motionless absorber, the output waveforms of the in-resonance photons and far-off-resonance photons are dramatically different (solid green and dotted black curves, respectively, in Fig.5b). In the case of vibrating absorber under the implemented AIT, they almost coincide (solid red and dotted black curves, respectively, in Fig.5c) showing that the waveform of the incident photon is preserved. The analysis of both spectral and temporal characteristics of the transmitted photons (Figs.4,5) shows also that AIT is a robust effect, which appearance does not require a strict compliance with the AIT conditions (1),(2).

In conclusion, the opaque two-level nuclear medium was made transparent for resonant γ-ray photons at room temperature by means of acoustic vibration of the sample. The proposed technique of AIT can be used for slowing down the γ-ray single-photon wave packet (which is expected to be more efficient compared to other techniques [10,11,31]) as well as for development of acoustically controllable interface between γ-ray photons and nuclear ensembles.

**ACKNOWLEDGEMENTS**

We are grateful to T. R. Akhmedzhanov for assistance in conducting experimental measurements and to Yury Shvyd'ko for the stimulating discussions. We acknowledge support by Russian Foundation for Basic Research (RFBR, Grants No. 18-32-00774 and No. 19-02-00852), as well as support by National Science Foundation (NSF, Grant No. PHY-150-64-67) and by AFOSR (Grant No. FA9550-18-1-0141) The numerical studies were supported by the Ministry of Science and Higher Education of the Russian Federation under Contract No. 14.W03.31.0032. I.R.Kh. acknowledges support by the Foundation for the Advancement of Theoretical Physics and Mathematics "BASIS". Y.V.R. acknowledges financial support (analytical studies) from the Government of the Russian Federation (Mega-Grant No. 14.W03.31.0028).